\documentclass[11pt]{cip-submit}
\usepackage{afterpage}
\usepackage{citesort}
\usepackage{float}
\usepackage{hyperref}
\usepackage{graphicx}
\usepackage{caption}
\usepackage{subcaption}
\usepackage{amsmath,amsxtra,amssymb,latexsym,amscd,color,bm}
\usepackage[mathscr]{eucal}
\usepackage{ulem}
\usepackage{cite}

\def\vsm{\vskip0.1cm}
\def\titles#1{\title{\large\bf\noindent #1}}
\def\authors#1{\author{\begin{flushleft}{#1}\end{flushleft}}}
\def\authord#1#2{\indent#1 \\
\textit{\indent#2}\vsm}

\def\email#1{\bigskip\href{mailto:#1}{\textit{E-mail:}~{#1}}\\[3mm]}

\def\Keywords#1{\\[.2cm] {\rm Keywords:~{#1}.}}
\def\Classification#1{\\[.2cm]{\rm Classification numbers:~{#1}.}}
\def\AND{$\text{\Small AND }$}
\def\and{$\text{\tiny AND }$}

\begin{document}
	\Year{2019}
	\Page{1}\Endpage{10}
\titles{Quantum transport through a ``charge" Kondo circuit:
effects of weak repulsive interaction in Luttinger Liquid}
\authors{\authord{T. K. T. Nguyen$^{1,\dagger}$ \AND M. N. Kiselev$^{2}$}{$^1$Institute of Physics, Vietnam Academy of Science and Technology, 10 Dao Tan, Ba Dinh, Hanoi, Vietnam\\
$^2$The Abdus Salam International Centre for Theoretical Physics,
Strada Costiera 11, I-34151, Trieste, Italy}
		$^\dagger$\email{nkthanh@iop.vast.ac.vn}
}

\begin{abstract}{\color{black}
We investigate theoretically quantum transport through the ``charge" Kondo circuit consisting of the quantum dot (QD) coupled weakly to an electrode at temperature $T+\Delta T$ and connected strongly to another electrode at the reference temperature $T$ by a single-mode quantum point contact (QPC). To account for the effects of Coulomb interaction in the QD-QPC setup operating in the integer quantum Hall regime we describe the edge current in the quantum circuit by Luttinger model characterized by the Luttinger parameter $g$. It is shown that the temperature dependence of both electric conductance $G\propto T^{2/g}$ and thermoelectric coefficient $G_T\propto T^{1+2/g}$ detours from the Fermi-liquid (FL) theory predictions. The behaviour of the thermoelectric power $S=G_T/G\propto T$ in a regime of a single-channel Kondo effect is, by contrast, consistent with the FL paradigm.
We demonstrate that the interplay between the mesoscopic Coulomb blockade in QD and weak repulsive interaction in the Luttinger Liquid $g=1-\alpha$ $(\alpha \ll 1)$ results in the enhancement of the thermopower. This enhancement is attributed to suppression of the Kondo correlations in the ``charge" circuit by the destructive quantum interference effects.}
\vspace{0.5cm}
\Classification{73.23.Hk, 73.50.Lw, 72.15.Qm, 73.21.La}
\vspace{0.5cm}
\end{abstract}

\maketitle
\markboth{}{}

\section{INTRODUCTION}
Quantum thermoelectricity is one of the rapidly developing topics of modern physics \cite{Whitney, Zlatic}. The search for new materials with enhanced thermoelectric properties continues to be a challenge for both theorists and experimentalists. For this purpose, quantum dot (QD) devices play significant role\cite{Blanter,kisbook}. The QD devices are  highly controllable and fine-tunable setups with adjustable external parameters such as bias voltage and magnetic field operating in- and out-of equilibrium.

In the QD-based quantum simulators, the QDs are typically connected to the leads (electric contacts) either by tunnel barriers or by quantum point contacts (QPCs). 
High tunability of the quantum simulators provides an access to fully control the dot-lead coupling whose strength is variable from weak to strong and vice versa. Based on this, the strong coupling regime, where the Kondo physics is important, can be easily achieved. The QD devices are therefore perfect playgrounds for the transport measurements including but not limited by investigation of the charge, spin and heat propagation in the quantum regime \cite{Beenakker, MA_theory,current_heat,thermo_exp2}.

\begin{figure}[t]
\centering
\includegraphics[width=85mm,angle=0]{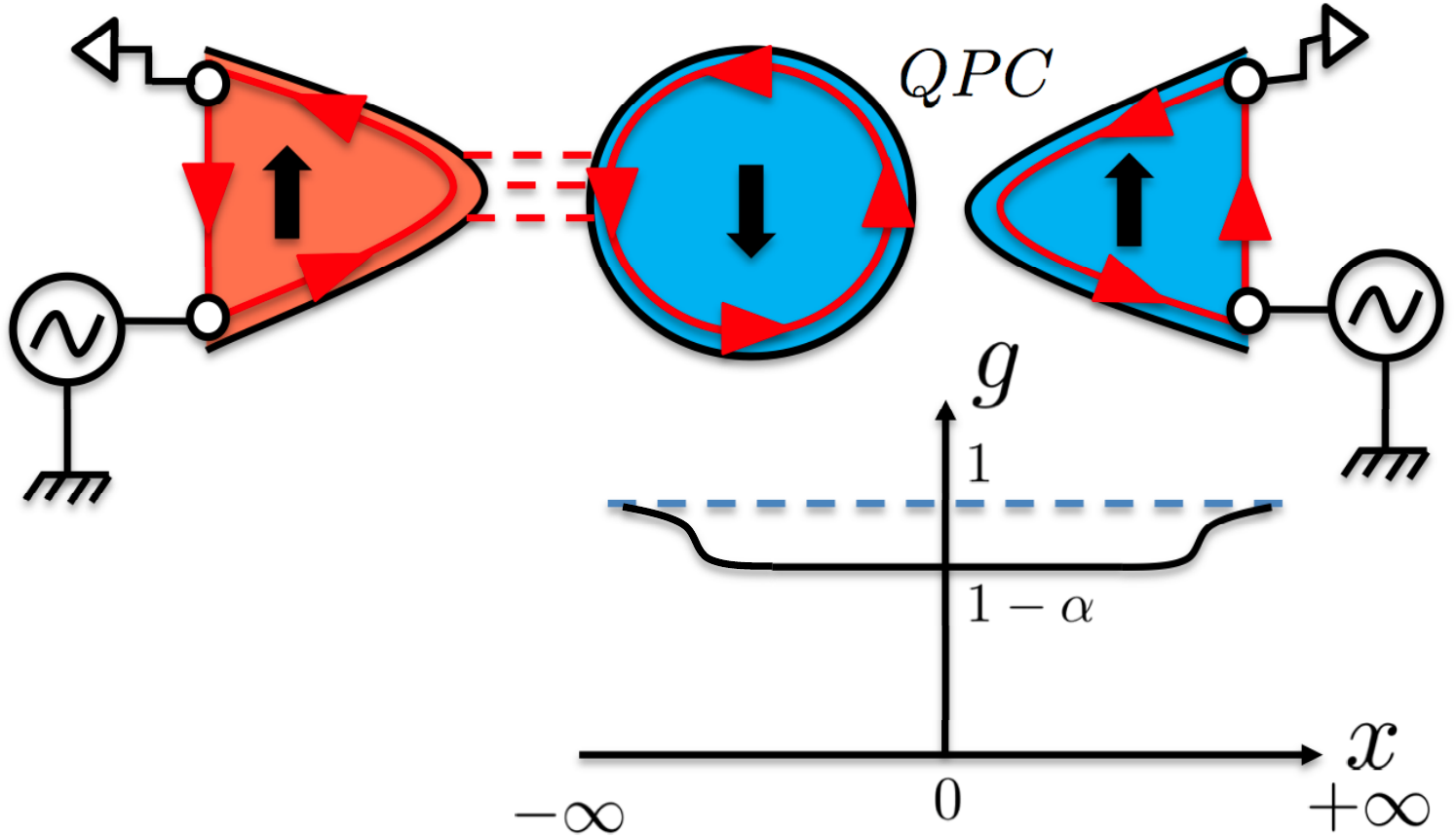} 
\vspace*{-0.1cm}
\caption{(Color online) Setup of a metal single electron transistor quantum device: the quantum dot (QD) is weakly coupled to the left electrode through a weak barrier and strongly coupled to the right electrode through a single mode quantum point contact (QPC). The QD and electrodes are formed by two-dimensional electron gas in the integer quantum Hall regime with $\nu=2$. The inner edge current is fully reflected (not shown) and the outer edge current is partially transmitted (solid red lines with arrows). The dark orange color stands for the higher temperature $T+\Delta T$ compared to the reference temperature $T$ of the blue pattern electrode. {\color{black} The lower panel shows the evolution of the interaction strength: the repulsive Coulomb interaction asymptotically vanishes ($g=1$) both at the position of the tunnel junction ($x\to -\infty$) and away from the QPC ($x\to +\infty$). The interaction adiabatically increases to a constant value $g=1-\alpha$, with $\alpha\ll 1$ in the QD-QPC area. Geometric size of QPC $L_{\rm QPC}$ is assumed to be smaller compared to characteristic length scale of the Coulomb potential variation.}}
\label{fig1}
\end{figure}

The theory of conventional Kondo effect has been developed and understood more than fifty years ago \cite{Kondo_1964}. Nevetheless, it still keeps attention of both theoretical and experimental communities. {\color{black} The observation of Kondo effect nowdays is confirmed} not only in the macroscopic bulk materials but also in micro- and meso-scale devices. The conventional Kondo effect is related to the spin degree of freedom of the quantum impurity \cite{Hewson}. However, new experiments in nano-structures operating in the integer quantum Hall (IQH) regime \cite{2CK_experiment_nature2015,3CK_experiment_science2018} in which two degenerate macroscopic charge states of a metallic
island (QD) play the role of iso-spin, convincingly prove an evidence for an unconventional ``charge" Kondo effect. The operational mechanism of the experimental ``charge" Kondo setups and physics of the quantum transport through the circuits is fully {\color{black} described by} pioneering theory proposed by Flensberg-Matveev-Furusaki \cite{Flensberg,Matveev1995,Furusaki_Matveev}. We refer to Ref. \cite{nk2018} for the detailed explanation {\color{black} of the connections between the ``charge" Kondo effect in recent quantum simulator experiments \cite{2CK_experiment_nature2015,3CK_experiment_science2018} and Flensberg-Matveev-Furusaki theory \cite{Flensberg,Matveev1995,Furusaki_Matveev,nk2018,LeHur,LeHur_Seelig,nkk_2010,nk2015}}.
In brief, the very simple idea of experimental realization of a single- and multi-channel Kondo simulators is as follows: the number of Kondo channels in the quantum device (see a cartoon on Fig.\ref{fig1}) is determined by the number of strong coupling QPCs connecting electrodes to QD. The electrons' location (inside the QD: iso-spin down, outside of QD: iso-spin up, see Fig.\ref{fig1}) plays the role of iso-spin flip processes. 

The physical observables in Kondo circuit setups are {\color{black} described} by the Fermi liquid (FL) theory \cite{Landau} in {\color{black} the following} situations: either the setup operates in a regime of a single QPC-QD (see Fig.\ref{fig1}),
or {\color{black} there is some controllable asymmetry between couplings of the QPCs and the QD in the multi-QPC setups} \cite{nkk_2010}. By contrast, the symmetric multi-QPCs devices are described by symmetric multi-channel Kondo models. These models are known to possess pronounced non-Fermi liquid properties \cite{Nozieres_Blandin_1980}. 

The motivation of this work is to consider the one-channel ``charge" Kondo (1CK) effect in a weakly Coulomb blockaded QD (see Fig.\ref{fig1}) with additional taking into account the effects of weak Coulomb interaction in the circuit. Most of the quantum transport observables for this model in the absence of Coulomb interaction fall to the Fermi liquid universality class. We postpone discussions of the non-Fermi liquid physics of the multi-QPC ``charge" Kondo circuits for future publications.

FL paradigm is known as one of the most important tools to describe the effects of interactions for most three- and two- dimensional electron systems \cite{Landau} ranging from  conventional metals to the heavy-electron materials \cite{Hewson}. However, it fails {\color{black} to describe the physics of} the interacting fermions in one dimension \cite{book_bosonization_1998,giamarchibook}. In this situation,the bosonization approach --  a bose representation of the fermion field -- is considered as a powerful technique. After bosonization, the Hamiltonian describing the fermion interactions is mapped into the noninteracting boson Hamiltonian, which forms the so-called Tomonaga-Luttinger (TL) model \cite{giamarchibook,Tomonaga,Luttinger,FL_LL1,FL_LL2}.

In the full-fledged TL model describing spinful electrons in 1D, the charge and spin sectors are disentangled. This phenomenon is known as a spin-charge separation. In this work, we investigate the quantum transport through a quantum simulator containing a quantum dot and a single short quantum wire (see Fig.\ref{fig1})
operating in the integer quantum Hall regime. Due to a strong quantized external magnetic field the electrons are spin-polarized (only one component of spin contributes to the scattering at the QPC, they can be considered as spinless electrons) and there is no any additional quantum number to be disentangled from the charge.  We describe the ``charge" Kondo circuit with a single conducting channel by the spinless TL model assuming that only charge mode is contributing to the quantum transport. The quantum transport coefficients of a 1CK model are calculated perturbatively with the effects of weak Coulomb repulsive interactions of the electrons included in the TL model. We discuss the scaling behaviour of the transport coefficients at the low temperature regime.  

The present paper is organized as follows. In Sec. II we describe the experimental setup and theoretical model as well as definitions of thermoelectric coefficients. The perturbatively analytic calculations and results are presented in Sec. III. Finally, in Sec. IV {\color{black} we present} the Conclusions.

\section{MODEL AND DEFINITIONS FOR QUANTUM TRANSPORT COEFFICIENTS}
We consider a setup (see Fig.\ref{fig1}) consisting of a large metallic QD with continuous spectrum. The source comprises the
QD being weakly coupled to the left lead through the tunnel barrier
with low transparency $|t|$$\ll $$1$ as described by a Hamiltonian $H_{tun}=\sum_{k}\left(t c_{k}^{\dagger}d+\text{h.c.}\right)$
where $c$ denotes the electrons in the left
lead, $d$ stands for the electrons in the dot. 
The temperature of the source can be controlled by the ``floating island"
technique \cite{quant_1}. The
temperature difference $\Delta T$ across the tunnel barrier is assumed
to be small compared to the reference temperature $T$ to guarantee
the linear response regime for the device at the weak link \cite{nkk_2010}. 
{\color{black} We account for the tunneling effects in the lowest order in $|t|^2$ to derive} the {\color{black} equations} for the thermo-
coefficients
which are related to the Matsubara Green's function (GF) of the electrons
in the dot at the left tunnel barrier 
\begin{eqnarray}
\mathcal G(\tau)=-\langle T_{\tau}d(\tau)d^\dagger(0)\rangle
=-\langle T_{\tau}\psi^{(0)}_\downarrow(\tau)F(\tau)F^\dagger(0)\psi_{\downarrow}^{\dagger}(0)\rangle,
\end{eqnarray}
in which $d(\tau)=\psi^{(0)}_\downarrow(\tau)F(\tau)$, $\psi^{(0)}_{\downarrow}$ is the fermionic operator describing electrons in the dot at the left weak link (see Fig. \ref{fig1}), $F$ is the operator lowering the integer-value operator of the number of electrons that entered the dot through the left weak link by unity (for more detailed explanation, see Refs. \cite{MA_theory, Furusaki_Matveev}). Since the operators $\psi^{(0)}_{\downarrow}$ and $F$ are decoupled, the Green function is factorized as $\mathcal G(\tau)=G_0(\tau)K(\tau)$, with $G_0(\tau)=\langle T_{\tau}\psi^{(0)}_\downarrow(\tau)\psi_{\downarrow}^{(0)\dagger}(0)\rangle$ {\color{black} being bare (non-interacting) GF}  and $K(\tau)=\langle T_\tau F(\tau)F^\dagger(0)\rangle$.

The drain comprises the QD electrically connected to a large electrode through a QPC at the reference temperature $T$. The QD and electrodes are formed by two-dimensional electron gas (2DEG) which is in the Integer Quantum Hall (IQH) regime at the filling factor $\nu=2$. The current propagating along the inner chiral edge channel
is fully reflected and can be ignored (not shown in Fig.\ref{fig1})
while the current propagating along the outer chiral edge channel (the red line with the direction shown by the red arrow)
is partially transmitted across the QPC. The mapping of
IQH setup to a (multi-channel) Kondo problem is explained in details in Ref. \cite{nk2018}. We notice that, the left QPC (the source) is tuned to weak coupling regime. Therefore, only the right strong coupling QPC (the drain) is considered as a quantum impurity in the Kondo phenomena. 
We assign the iso-spin $\uparrow$, $\downarrow$ to the electrons in the QPC and QD correspondingly. 
The ``charge" iso-spin flips when the electrons move in- and out- of the QD. Backscattering transfers ``moving in-" the QD electrons to ``moving out-" from the QD electrons and vice versa.
The single QPC is equivalent to the single channel in the $S=1/2$ Kondo problem \cite{Hewson} which corresponds to the spinless case in the Andreev-Matveev theory \cite{MA_theory}. 

In the spirits of Andreev-Matveev theory \cite{MA_theory}, after some straightforward manipulations we come to the {\color{black} Euclidean} action for
the Luttinger model $S=S_{0}+S_{C}+S^{\prime}$ at the drain with 
\begin{equation}
S_{0}=\frac{1}{2\pi g}\!\int_{0}^{\beta}\!\!\!dt\int_{-\infty}^{\infty}\!\!\!\!\!dx\!\!\left[\frac{\left(\partial_{t}\phi(x,t)\right)^{2}}{u}+u\left(\partial_{x}\phi(x,t)\right)^{2}\right]\label{eq:S0_LM_1CK}
\end{equation}
describing the electron transport through the right QPC, with {\color{black} $\phi(x,t)$ are bosonic fields, the alter-ego of 1D fermions,} $u$ is the charge mode renormalized velocity and $g$ is the charge mode Luttinger parameter attributed to repulsive Coulomb interaction at the edge, {\color{black} $\beta=1/T$ (we adopt the units $\hbar$$=$$c$$=$$k_B$$=$$1$ in all equations)}. {\color{black} The exact value of the interaction parameter $g$ depends on the microscopic model. However, without loss of generality it can be estimated as
\begin{equation}
g\approx \left[1+U/(2\epsilon_F)\right]^{-1/2},
\label{g_def}
\end{equation}
where $U\sim e^2/(\varepsilon a)$ is Coulomb interaction, $a$ is a distance between electrons, $\varepsilon$ is a dielectric constant and $\epsilon_F$ is a Fermi energy in the 2DEG \cite{GRS,Kane_Fisher}.}
We notice that in this work, we consider the regime $g=1-\alpha$ with $0<\alpha\ll 1$,  consistent with the regime of Luttinger parameters investigated in the experiments  \cite{LLparameter_exp1,LLparameter_exp2}. In fact, from the theory of Luttinger liquid, this regime corresponds to the situation that the scattering of left/right onto right/left electrons (dispersion scattering) is repulsive and weak. {\color{black} The treatment of the weak repulsive interacting regime characterized by the Luttinger parameter $g$ is justified through the validity of perturbative calculations (c.f. \cite{Kane_Fisher}). We discuss the low bound for $g$ in the Section III.}  The action $S_{C}$ 
\begin{equation}
S_{C}\left(\tau\right)=\int_{0}^{\beta}\!\!dtE_{C}\left[n_\tau(t)+\frac{1}{\pi}\phi(0,t)-N(V_{g})\right]^{2}\label{eq:Sc_1CK}
\end{equation}
describes the Coulomb interaction in the dot {\color{black} where the number of electrons $N(V_g)$ is controlled by the gate voltage $V_g$ and $E_C$ is a charging energy accounting for the effects of weak charge quantization in QD known as mesoscopic Coulomb Blockade \cite{aleiner_glazman}.} {\color{black} Finally,}  
\begin{eqnarray}
S^{\prime} & = & -\frac{D}{\pi}|r|\int_{0}^{\beta} dt\cos\left[2\phi(0,t)\right].\label{eq:Sst_1CK}
\end{eqnarray}
{\color{black} represents the backscattering at the right QPC. Here $|r|$ is a reflection amplitude associated with the backscattering processes and $D$$\sim $$\epsilon_F$ is an ultra-violet cutoff of the theory (bandwidth of the 2DEG).}

Note, that the operator $\psi_{\downarrow}(-\infty)$ {\color{black} used in the definition of the GF}
is expressed through the fermionic operators $\psi_{\downarrow}(x){\sim}e^{i\phi(x)}$
in the one dimensional channel describing the chiral edge
QD using the standard bosonization technique \cite{book_bosonization_1998}
(see also \cite{MA_theory} and \cite{nkk_2010} for more details).
The scattered states of the right lead-QPC are
bosonized as $\psi_{\uparrow}(x){\sim}e^{-i\phi(x)}$. The function $n_\tau(t)$ in the charging action $S_{C}$ accounts for the electrons entering the dot through the left weak tunnel barrier. The number of electrons increases from $0$ to $1$ at time $t=0$ and decreases from $1$ back to $0$ at time $t=\tau$. Therefore, $n_{\tau}(t)=\theta(t)\theta(\tau-t)$. Here $\theta(t)$ is the unit step function.

The Luttinger parameter is assumed $g=1$ (see Fig.\ref{fig1}) at the left contact ($x=-\infty$), therefore, the Green's function $G_0(\tau)$ at the left weak link can be expressed as    
\begin{equation}
G_0\left(\tau\right)=-\frac{\nu_0\pi T}{\sin\left(\pi T\tau\right)},
\end{equation}
with $\nu_0$ is the density of states in the dot without interactions. The correlation function $K(\tau)$ can be calculated as
\begin{eqnarray}
K(\tau) & = & Z(\tau)/Z(0),\nonumber \\
Z(\tau) & = & \int\exp[-S_{0}-S_{C}(\tau)-S^{\prime}]\prod_{\alpha}\mathcal{D}\phi_{\alpha}(x,t)\label{zz}
\end{eqnarray}

The electric conductance \cite{Furusaki_Matveev} is given by 
\begin{eqnarray}
G & = & \frac{G_{L}\pi T}{2}\int_{-\infty}^{\infty}\frac{1}{\cosh^{2}(\pi Tt)}K\left(\frac{1}{2T}+it\right)dt~.\label{elec_cond_def}
\end{eqnarray}
Here $G_{L}\ll e^{2}/h$ denotes the tunnel conductance of the left
barrier calculated ignoring influence of the dot. The thermal conductance
takes the form \cite{MA_theory} 
\begin{eqnarray}
G_{T} & = &-\frac{i\pi^{2}}{2}\frac{G_{L}T}{e}\int_{-\infty}^{\infty}\frac{\sinh(\pi Tt)}{\cosh^{3}(\pi Tt)}K\left(\frac{1}{2T}+it\right)dt~.\label{thercond_def}
\end{eqnarray}
The Seebeck effect quantified in terms of the thermoelectric power
(TP) {\color{black} for the zero-current state when the electric current associated with the temperature drop $\Delta T$ is nullified by applying a thermo-voltage 
$\Delta{\mathcal V}_{{\rm th}}$}. Thermopower $S$ is expressed as a ratio of the transport coefficients $G$ and $G_{T}$ in linear response regime as: 
\begin{equation}
S=-\left.\frac{\Delta{\mathcal V}_{{\rm th}}}{\Delta T}\right|_{I_{{\rm sd}}=0}=\frac{G_{T}}{G}.
\label{eq:TP}
\end{equation}

\section{PERTURBATIVE RESULTS}
\label{Sec4}

We perform perturbative calculations respecting to the backscattering
for each of the two above models. These calculations are in the spirits
of Matveev-Andreev theory \cite{MA_theory}, which concern the saddle-point method.

We first evaluate Gaussian integral in Eq. (\ref{zz}) for the action
of the one channel Kondo model as shown in Eqs. (\ref{eq:S0_LM_1CK}, \ref{eq:Sc_1CK}, \ref{eq:Sst_1CK})
under the assumption $S^{\prime}=0$. At zero order, the saddle point,
based on the principle of the action minimum, is found as 
\begin{equation}
\phi_{\tau}\left(x,t\right)=\pi N-T\sum_{\omega_{n}}\frac{gE_{C}\exp\left[-\frac{|\omega_{n}x|}{u}\right]}{|\omega_{n}|+\frac{gE_{C}}{\pi}}n_{\tau}\left(\omega_{n}\right)e^{-i\omega_{n}t},
\label{saddle_point}
\end{equation}
with $\omega_{n}=2\pi nT$ are bosonic Matsubara frequencies and the
Fourier transform of $n_{\tau}(t)$ is $n_{\tau}\left(\omega_{n}\right)=\left(e^{i\omega_{n}{\color{black} \tau}}-1\right)/i\omega_{n}$.

In the calculation $K_{0}(\tau)=Z(\tau)/Z(0)$, the integrals over
the fluctuations of the field $\phi\left(x,t\right)$ {\color{black} around} the saddle
points in the numerator and the denominator cancel each other. Therefore,
the value of $K_{0}(\tau)$ is evaluated by the integrals at the saddle
point values. In the condition $\tau\gg E_{C}^{-1}$ and $T\ll E_{C}$,
we find 
\begin{equation}
\left[S_{0}+S_{C}(\tau)\right]_{\phi=\phi_{\tau}\left(x,t\right)}=\frac{E_{C}}{\pi^{2}T}\sum_{n=1}^{\infty}\frac{\left[1-\cos\left(2\pi Tn\tau\right)\right]}{n\left[n+\frac{gE_{C}}{2\pi^{2}T}\right]},
\end{equation}
and $\left[S_{0}+S_{C}(\tau)\right]_{\phi=\phi_{0}\left(x,t\right)}=0$.
The correlator $K_{0}(\tau)$
\begin{equation}
K_{0}\left(\tau\right)=\left[\frac{\pi^{2}T}{g\gamma E_{C}}\frac{1}{|\sin\left(\pi T\tau\right)|}\right]^{2/g}.\label{eq:KC}
\end{equation}
{\color{black} Here $\gamma$$=$$e^{\bf C}$$\approx $$1.78$, where ${\bf C}$$\approx $$0.577$ is Euler's constant.}
The correlation function {\color{black} $K_{0}\left(\tau\right)$ is computed with the Gaussian action $S_{0}+S_{C}(\tau)$ and therefore corresponds to the particle-hole symmetric case. The particle-hole symmetry is broken by the backscattering on the QPC considered perturbatively below.}
Plugging formula (\ref{eq:KC}) into formula (\ref{elec_cond_def}) we
find the electric conductance as a function of the temperature as
\begin{eqnarray}
G & = & \frac{G_{L}}{2}C\left(g\right)\left[\frac{T}{E_{C}}\right]^{\frac{2}{g}}~,
\label{electric_cond1}
\end{eqnarray}
{\color{black} where} 
\begin{equation}
C\left(g\right)=\left[\frac{\pi^{2}}{g\gamma}\right]^{\frac{2}{g}}\int_{-\infty}^{\infty}\frac{1}{\cosh^{2+\frac{2}{g}}\left(x\right)}dx
\label{constantC}
\end{equation}
{\color{black} is a temperature-independent constant which depends only on the value of the Luttinger parameter $g$ (c.f. this result with corresponding equations of Kane-Fisher theory \cite{Kane_Fisher}).}

Plugging formula (\ref{eq:KC}) into formula (\ref{thercond_def}) we
find that the thermal conductance vanishes. We thus need to calculate
the first order of perturbation theory respecting the backscattering
amplitude $|r|$:
\begin{equation}
K(\tau)=K_0(\tau)\left(1-\langle S^\prime\rangle_\tau+\langle S^\prime\rangle_0\right).
\end{equation}
This correlation function is characterized by the following symmetries
associated with particle-hole symmetry and shift transformation: $K(\beta-\tau,N)=K(\tau,1-N)$ and $K(\beta-\tau,N)=K(\tau,-N)$. The average of the backscattering action can be calculated through the averaging over the fluctuations $\varphi=\phi-\phi_\tau$ around the saddle point $\phi_\tau(x,t)$:
\begin{eqnarray}
\langle S^{\prime}\rangle_{\tau}&=&-\frac{D}{\pi}|r|{\rm Re}\left[ \int^{\beta}_{0} dt e^{2i\phi_\tau(0,t)}\left\langle e^{2i\varphi(0,t)}\right\rangle\right]\nonumber\\
&=&-\frac{\gamma}{\pi^2}|r|gE_C{\rm Re}\left[\int^{\beta}_{0}dt e^{2i\phi_\tau(0,t)}\right], 
\label{s'}
\end{eqnarray}
Plugging formula (\ref{saddle_point}) into formula (\ref{s'}) we
find
\begin{equation}
\langle S^{\prime}\rangle_{\tau}-\langle S^{\prime}\rangle_{0}=-\frac{\gamma}{\pi^{2}}|r|gE_{C}{\rm Re}\left[e^{i2\pi N}\int_{0}^{\beta}dt\left[e^{i\left[F\left(t\right)-F\left(t-\tau\right)\right]}-1\right]\right],
\end{equation}
where 
\begin{equation}
F\left(t\right)=2\sum_{n=1}^{\infty}\frac{\sin\left(2\pi Ttn\right)}{n+\frac{gE_{C}}{2\pi^{2}T}}.
\end{equation}
At $T\ll E_{C}$ and $K\left(\tau\right)$ is considered in the regime
$\tau\sim T^{-1}\gg E_{C}^{-1}$ we find 
\begin{equation}
\langle S^{\prime}\rangle_{\tau}-\langle S^{\prime}\rangle_{0}=2\gamma\xi|r|\left[\cos\left(2\pi N\right)-\frac{2\pi^{2}T}{gE_{C}}\sin\left(2\pi N\right)\cot\left(\pi T\tau\right)\right],
\end{equation}
with $\xi\approx 1.59$. The thermoelectric coefficient $G_T$ must be an odd function of the gate voltage $N$ {\color{black} (we use a shorthand notation $N$$\equiv $$N(V_g)$)} as well as odd function of the imaginary time $\tau$ \cite{MA_theory}. Therefore, the result for thermoelectric coefficient at the first order of the perturbation in $|r|$ is
\begin{equation}
G_{T}=-2\frac{G_{L}}{e}\pi^{3}\gamma\xi|r|C_{T}\left(g\right)\left[\frac{T}{E_{C}}\right]^{1+\frac{2}{g}}\sin\left(2\pi N\right),
\label{thercond_result}
\end{equation}
{\color{black} (we refer again to the results of the Kane-Fisher theory \cite{Kane_Fisher_TP} for comparison), and}
\begin{equation}
C_{T}\left(g\right)=\frac{1}{g}\left[\frac{\pi^{2}}{g\gamma}\right]^{\frac{2}{g}}\int_{-\infty}^{\infty}\frac{\sinh^{2}\left(x\right)}{\cosh^{4+\frac{2}{g}}\left(x\right)}dx.
\label{constantCT}
\end{equation}
Substituting Eq. (\ref{electric_cond1}) and Eq. (\ref{thercond_result})
into Eq. (\ref{eq:TP}), we obtain the expression for the TP as 
{\color{black}
\begin{equation}
S=-\frac{4\pi^{3}\gamma\xi}{5 e}|r|C_S\left(g\right)\frac{T}{E_{C}}\sin\left(2\pi N\right),
\label{TP_result}
\end{equation}
}
with 
{\color{black}
\begin{equation}
{\color{black} C_S}\left(g\right)\equiv 5 \frac{C_{T}\left(g\right)}{C\left(g\right)}=\frac{5}{g}\left[\int_{-\infty}^{\infty}\frac{\sinh^{2}\left(x\right)}{\cosh^{4+\frac{2}{g}}\left(x\right)}dx\right]\left[\int_{-\infty}^{\infty}\frac{1}{\cosh^{2+\frac{2}{g}}\left(x\right)}dx\right]^{-1}.
\label{tildeC}
\end{equation}
}
  \begin{figure}
 	\centering
 	\includegraphics[width=10cm]{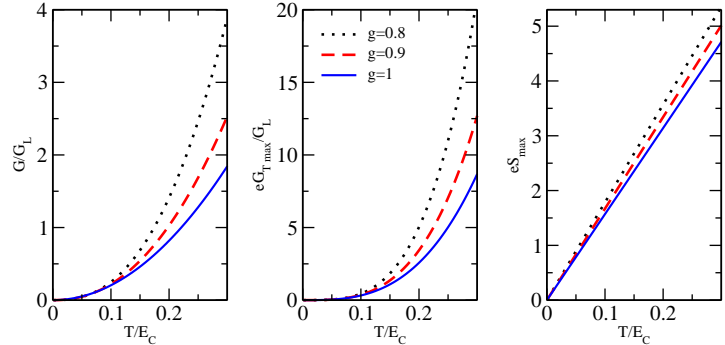}
 	\vspace*{-25mm}
 	\caption{(Color online) Plots of electric conductance $G/G_L$ (left panel), maximum of thermal coefficient $eG_{T\; max}/G_L$ (center panel), and maximum of thermopower $eS_{max}$ (right panel) as functions of temperature $T/E_C$ for different Luttinger parameter values $g=0.8$ (dotted black lines), $g=0.9$ (dashed red lines), and $g=1$ (solid blue lines) {\color{black} at} $|r|^2=0.05$.} 
 	\label{fig2}
 \end{figure}
The Eq. (\ref{TP_result}) is the {\color{black} central} result of this paper. The TP of the 1CK model with interactions in the Luttinger liquid depends linearly on temperature $T$, following the FL picture similar to the result of the non-interacting electron system as shown in the Eq. (35) of Ref.\cite{MA_theory}. The effects of interaction are incorporated into the pre-factor {\color{black} $C_S$}. As we consider the system in the regime $g=1-\alpha$ with $\alpha \ll 1$, we expand the integrals in formulas (\ref{constantC}) and (\ref{constantCT}) in series of $\alpha$ and stop 
{\color{black} expansion} at the first order. {\color{black} As a result,} we obtain
{\color{black}
\begin{equation}
{\color{black} C_S}\left(g=1-\alpha\right)\approx\frac{1}{1-\alpha}\times \frac{1-\left(\frac{31}{15}-2\ln 2\right)\alpha}{1-\left(\frac{5}{3}-2\ln 2\right)\alpha}\approx
1+\frac{3}{5}\alpha+O(\alpha^2).
\label{tildeC1}
\end{equation}
}
The Eq. (\ref{tildeC1}) represents the weak dependence of TP on the interaction in the Luttinger liquid. {\color{black} In the non-interacting limit $g=1$ (or, equivalently, $\alpha=0$) Eq.(\ref{TP_result}) coincides with the result of Matveev-Andreev theory. The validity of the perturbation theory for $G_T$ is controlled by $|r|C_T(g)\ll 1$ which sets the low bound for the Luttinger parameter $g_{\rm min}<g\leq 1$.  The estimation for $g_{\rm min}$ follows from the connection between $g_{\rm min}$ and $|r|$: $|r|C_T(g_{\rm min})\sim 1$. For $|r|=0.1$ numerical value of the low bound $g_{\rm min}\approx 0.72$.}
 
The result in Eqs. (\ref{electric_cond1}, \ref{thercond_result}, \ref{TP_result}) are {\color{black} illustrated} in Fig. \ref{fig2}. The electric conductance $G/G_L$ (left panel), maximum of thermal coefficient $eG_{T\; max}/G_L$ (central panel), and maximum of TP $eS_{max}$ (right panel) are plotted as functions of temperature $T/E_C$ for different Luttinger parameter values $g$. We find that the thermoelectric coefficients are enhanced when $g$ is smaller but close to $1$. The enhancement of thermopower due to weak repulsive interaction can be explained as follows. The effects of Coulomb interaction in the QPC
(short quantum wire) suppress the mesoscopic
Coulomb blockade (reduce charging energy of the dot).
Since $E_C$ plays the role of Kondo temperature, we attribute enhancement of TP to the reduction of the Kondo correlations due to the destructive quantum interference effects.

\section{CONCLUSIONS}
In summary, we derived the scaling relations for the TP of the 1CK ``charge" Kondo model approaching the FL strong coupling fixed point with the effect of electron interaction in the Luttinger {\color{black} Liquid}, which describes the IQH edge current. The temperature dependence of the electric conductance, thermoelectric coefficient and TP are calculated perturbatively in the temperature regime {\color{black} $T\ll E_C$}. Although the electric conductance $G$ and the thermoelectric coefficient $G_T$ vanish as a power law with an exponent inversely proportional to the Luttinger parameter when the temperature goes down, the TP $S$ depends linearly on temperature. The Luttinger parameter dependence of the TP of 1CK model appearing in the pre-factor shows a slight deviation from corresponding behaviour of non-interacting 1CK model. Therefore, we conclude that the Fermi liquid temperature scaling of TP remains in a weak repulsive interaction regime of a Luttinger liquid model. This opens access to experimental measurements {\color{black} of the Luttinger interaction parameter} in the quantum thermoelectric transport experiments. 

\section*{ACKNOWLEDGEMENT}

This research in Hanoi is funded by Vietnam National Foundation for Science and Technology Development (NAFOSTED) under grant number 103.01-2016.34.

\end{document}